\documentclass[prx, notitlepage, twocolumn,nofootinbib,superscriptaddress,longbibliography]{revtex4-1}
\usepackage[colorlinks=true,citecolor=blue,linkcolor=blue]{hyperref}
\usepackage{graphicx}
\usepackage{color}
\usepackage{dcolumn}
\usepackage{bm}
\usepackage{mathrsfs}
\usepackage{amsmath}
\usepackage{amssymb}
\usepackage{amsthm}
\usepackage{amsfonts}
\usepackage{enumerate}
\usepackage{latexsym}
\usepackage{hyperref}
\usepackage{centernot}
\usepackage{multirow}

\begin{document}

\title{Calculated magnetic exchange interactions in Dirac magnon material Cu$%
_{3}$TeO$_{6}$}
\author{Di Wang}
\affiliation{National Laboratory of Solid State Microstructures and School of Physics,
Nanjing University, Nanjing 210093, China}
\affiliation{Collaborative Innovation Center of Advanced Microstructures, Nanjing 210093,
China}
\author{ Xiangyan Bo}
\affiliation{National Laboratory of Solid State Microstructures and School of Physics,
Nanjing University, Nanjing 210093, China}
\affiliation{Collaborative Innovation Center of Advanced Microstructures, Nanjing 210093,
China}
\author{Feng Tang}
\affiliation{National Laboratory of Solid State Microstructures and School of Physics,
Nanjing University, Nanjing 210093, China}
\affiliation{Collaborative Innovation Center of Advanced Microstructures, Nanjing 210093,
China}
\author{Xiangang Wan}
\affiliation{National Laboratory of Solid State Microstructures and School of Physics,
Nanjing University, Nanjing 210093, China}
\affiliation{Collaborative Innovation Center of Advanced Microstructures, Nanjing 210093,
China}
\thanks{Corresponding author: xgwan@nju.edu.cn}

\begin{abstract}
Recently topological aspects of magnon band structure have attracted much
interest, and especially, the Dirac magnons in Cu$_{3}$TeO$_{6}$ have been
observed experimentally. In this work, we calculate the magnetic exchange
interactions \textit{J}'s using the first-principles linear-response
approach and find that these $J^{\prime }s$ are short-range and negligible
for the Cu-Cu atomic pair apart by longer than 7 \AA . Moreover there are
only 5 sizable magnetic exchange interactions, and according to their signs
and strengths, modest magnetic frustration is expected. Based on the
obtained magnetic exchange couplings, we successfully reproduce the
experimental spin-wave dispersions. The calculated neutron scattering cross section also agrees very
well with the experiments. We also calculate Dzyaloshinskii-Moriya
interactions (DMIs) and estimate the canting angle ($\sim $ 1.3$%
{{}^\circ}%
$) of the magnetic non-collinearity based on the competition between DMIs
and \textit{J'}s, which is consistent with the experiment. The small canting
angle agrees with that the current experiments cannot distinguish the DMI
induced nodal line from a Dirac point in the spin-wave spectrum. Finally we
analytically prove that the ``sum rule'' conjectured in [Nat. Phys. \textbf{%
14}, 1011 (2018)] holds but only up to the 11th nearest neighbour.
\end{abstract}

\date{\today }
\maketitle

\affiliation{National Laboratory of Solid State Microstructures and School of Physics,
Nanjing University, Nanjing 210093, China}
\affiliation{Collaborative Innovation Center of Advanced Microstructures, Nanjing 210093,
China}

\affiliation{National Laboratory of Solid State Microstructures and School of Physics,
Nanjing University, Nanjing 210093, China}
\affiliation{Collaborative Innovation Center of Advanced Microstructures, Nanjing 210093,
China}

\affiliation{National Laboratory of Solid State Microstructures and School of Physics,
Nanjing University, Nanjing 210093, China}
\affiliation{Collaborative Innovation Center of Advanced Microstructures, Nanjing 210093,
China}

\affiliation{National Laboratory of Solid State Microstructures and School of Physics,
Nanjing University, Nanjing 210093, China}
\affiliation{Collaborative Innovation Center of Advanced Microstructures, Nanjing 210093,
China}

\affiliation{National Laboratory of Solid State Microstructures, Collaborative Innovation
Center of Advanced Microstructures and College of Physics, Nanjing
University, Nanjing 210093, China}

\section{Introduction}

The non-trivial topological nature of electronic bands has been studied
extensively during the past decade \cite{ti-1,ti-2}. Plenty of topological
materials have been discovered, such as topological insulator \cite%
{ti-1,ti-2}, Dirac semimetal \cite%
{dirac-1,dirac-2,dirac-3,dirac-4,dirac-5,dirac-6}, Weyl semimetal \cite%
{weyl-1,weyl-2,weyl-5} and node-line semimetal \cite%
{node-1,node-2,node-3,node-4,node-5}, etc \cite{Basil}. In addition to the
above topological phases, the rich variety of spatial symmetries in
condensed matter systems results in various novel topological crystalline
insulators/semimetals \cite{TCI-1,hourglass,np,TCI-4,Fang-5,TCI-6,nodalchain}%
. By exploiting the mismatch between the real and momentum-space
descriptions of the band structure, a complete classification scheme of band
topology has been proposed \cite{Po-1,QCT,2D-Symmetry,magnetic-Symmetry}. A
comprehensive database search for ideal non-magnetic\ topological materials
has been finished \cite{Tang-3} by combining first principles calculation
and the symmetry-indicator theory \cite{Tang-1}. Meanwhile, thousands of
electronic topological materials have also been proposed based on the graph
theory \cite{BAB} and the complete mapping between the symmetry
representation of occupied bands and the topological invariants \cite{IOP}.

It is worth mentioning that the topological feature is not only restricted
to electronic systems. The band crossings in systems of photons \cite%
{topophoton-1,topophoton-2,topophoton-3,topophoton-4} and phonons \cite%
{topophonon-1,topophonon-3} have also been intensively investigated.
Moreover,\ recent research in the magnon systems has leaded to the discovery
of topological magnon insulators \cite{topomagnonti-1,topomagnonti-2},
magnonic Dirac semimetals \cite%
{topomagnondirac-1,topomagnondirac-2,topomagnondirac-3} and Weyl semimetals
\cite{topomagnon-1,topomagnon-2,topomagnon-3}.

In 2017, Li et al. \cite{ref1} proposed Dirac magnons may occur in the
three-dimensional antiferromagnetic material Cu$_{3}$TeO$_{6}$. As shown in
Fig. \ref{cry}, Cu$_{3}$TeO$_{6}$ crystallizes in the centrosymmetric cubic
crystal structure (space group $Ia$-$3$) \cite{ctostr,ctostr2}. Temperature (%
$T$) dependent magnetic susceptibility ($\chi (T)$) reveals that this
compound displays a long-range antiferromagnetic (AFM) ordering below $T$$%
_{N}\sim 60\ $K \cite{ctoexp-2}. Within the range of $T=180-330$ K,\ $\chi
(T)$\ can be fitted very well by the Curie-Weiss (CW) law with the CW
temperature $\theta _{CW}=-130$ K \cite{ctoexp-2}. The $\chi (T)$\ deviates
from CW behavior below 180 K, which is three times larger than $T$$_{N}$.
This may indicate the frustrated magnetic feature \cite{ctoexp-2}. A clear
bulk magnetic transition at around 62 K has also been observed by muon-spin
relaxation/rotation measurement \cite{ctoexp}. Neutron powder diffraction
experiment \cite{ctoexp-2} suggests two possible magnetic configurations:
(i) collinear AFM order (ii) non-collinear configuration. In the collinear
case, the two spins connected by inversion ($\mathcal{P}$) have opposite
spin orientations \cite{ctoexp-2}, thus Cu$_{3}$TeO$_{6}$ is invariant under
$\mathcal{P}\mathcal{T}$ symmetry ( $\mathcal{T}$ is the time-reversal
transformation), protecting robust magnon Dirac points \cite{ref1}. For the
non-collinear magnetic case \cite{ctoexp-2}, Li et al. \cite{ref1} propose
that non-collinearity breaks the U(1) symmetry. As a result, the Dirac point
in the collinearly magnetically ordered state expands into a nodal line \cite%
{ref1}.

Motivated by this theoretical prediction, Yao et al. \cite{ref2} and Bao et
al. \cite{ref3} have measured spin excitations of Cu$_{3}$TeO$_{6}$ with
inelastic neutron scattering (INS), respectively. Both of them have observed
the existence of band crossing points in the magnon spectra \cite{ref2,ref3}%
. In addition to Dirac points, at $\Gamma $ and $H$ points of the Brillouin
zone (BZ) Bao et al. \cite{ref3} also observed the triply degenerate nodes
which can also occur in electronic bands \cite{new-fermion,MoP}. Bao et al.
\cite{ref3} found that the experimental magnon band dispersion can be well
reproduced by a spin Hamiltonian dominated by only the 1st nearest-neighbour
(NN) exchange interaction \textit{J}$_{1}$. While Yao et al. \cite{ref2}\
suggested that the magnetic moments in this compound couple over a variety
of distances, and even the ninth-nearest-neighbour \textit{J}$_{9}$ plays an
important role. Strikingly they found an interesting relation between magnon
eigenvalues at different high symmetry points of BZ which was dubbed as
``sum rule'' \cite{ref2}.\newline

Generally, spin-orbit coupling (SOC) always exists and leads to the
Dzyaloshinskii-Moriya interactions (DMIs) \cite{dm,dm2} even in the
centrosymmetric compound Cu$_{3}$TeO$_{6}$, as discussed in the following
sections. The DMIs could result in a non-collinearity in the ground magnetic
state, leading to nodal lines in magnetic excitations \cite{ref1}. As
mentioned above, the two experiments \cite{ref2,ref3} have observed the
existence of Dirac points\ but cannot identify the nodal lines from the band
crossing points. Note that the size of nodal lines strongly depends on the
canting angle of the non-collinearity, which is determined by the
competition between exchange interaction $J$ and DMI \cite{ref1}. Therefore
it is an interesting issue to obtain accurate spin exchange parameters $%
J^{\prime }s$ and DMIs, which we address in the current work.

In this paper, based on first-principles calculations, we systematically
study the electronic and magnetic properties of Cu$_{3}$TeO$_{6}$. The
calculations show that Cu$_{3}$TeO$_{6}$ is an insulator with a band gap
about 2.07 eV. The calculated magnetic moment of Cu ions is about 0.81 $%
\mu
_{B}$, which is larger than the experimental value (0.64 $%
\mu
_{B}$) measured by the neutron powder diffraction \cite{ctoexp-2}. Using a
first-principles linear-response (FPLR) approach \cite{ourJ1}, we calculate
the spin exchange parameters $J^{\prime }s$. Based on these spin exchange
parameters,\ we calculate the magnetic excitation spectra using linear spin
wave theory (LSWT) and the calculated spin wave spectra agree with the
experiments very well, as well as the positions of the Dirac and triply
degenerate magnons in the BZ \cite{ref2,ref3}. We also calculate the neutron
scattering cross section, which is consistent with the experiment \cite%
{ref2,ref3}. The calculated exchange interactions are short-range and
negligibly weak for the distance more than 7 \r{A}. There are only five
sizable magnetic exchange terms and all them favor antiferromagnetic
ordering. These spin exchange parameters are compatible with the modest
frustration in Cu$_{3}$TeO$_{6}$ according to their signs and magnitudes. We
also analytically prove that the magnon energies at high symmetry points of
BZ cannot own a general \textquotedblleft sum rule\textquotedblright\
conjectured in Ref. \cite{ref2} which is {found to be} only satisfied up to
the 11th NN. Moreover, we also calculate the DMIs and estimate the canting
angle of non-collinearity which is about 1.3$%
{{}^\circ}%
$, consistent with the experimental value $\sim $ 6%
${{}^\circ}$
\cite{ctoexp-2}. This may be the reason why the recent experimental works
only observed the existence of Dirac points\ instead of the nodal lines \cite%
{ref2,ref3}.

\section{Method}

The electronic band structure and density of states calculations are carried
out by using the full potential linearized augmented plane wave method as
implemented in WIEN2K package \cite{wien2k}. Local spin density
approximation (LSDA) for the exchange-correlation potential is used here. A
10$\times $10$\times $10 k-point mesh is used for the Brillouin zone
integral. Using the second-order variational procedure, we include the SOC
interaction \cite{socref}. The self-consistent calculations are considered
to be converged when the difference of the total energy of the crystal does
not exceed 0.01 mRy. We utilize the LSDA + $U$ scheme \cite{LDA+U} to take
into account the effect of Coulomb repulsion in {Cu-}$3d$ orbital. The value
of $U=10$ eV and $J=1$ eV for Cu-oxides works well in the previous
theoretical work \cite{u10,ourJ2}. We vary the parameter $U$\ between 8.0
and 10.0 eV and find that our results are not sensitive to the values of $U$
in this range. Thus in this paper we show our results for $U=10$ eV.

The spin exchange parameters $J^{\prime }s$\ are the basis to understand
magnetic properties. By fitting $J$ to reproduce experimental results, such
as $\chi (T)$\ and magnon dispersion, one can extract the exchange
interaction parameters $J^{\prime }s$\. However, an unambiguous fitting is basically
impossible. For example, as mentioned above, though the INS experimental
results {of Yao et al. and Bao et al.} are consistent with each other, their
fitting results of the spin exchange interactions are completely different
\cite{ref2,ref3}. In addition to this phenomenological approach, theoretical
calculations can also be used to estimate the exchange interaction
parameters. A popular numerical method to calculate $J$ is to calculate the
total energies of the $N+1$ magnetic configurations, and map it by a spin
Hamiltonian to extract $N$ exchange constants. Unfortunately this
theoretical method has several drawbacks: (i) the calculated magnetic
moments may depend on the magnetic ordering, which significantly affect the
accuracy of the obtained \textit{J}; (ii) it is not clear that how many
exchange interactions \textit{J} one need to use when mapping the total
energies from the first-principles calculation on the spin Hamiltonian. An
alternative but much more efficient method to calculate spin exchange
interactions by first-principles is based on combining magnetic force
theorem and linear-response\ approach \cite{Jref1}. The exchange interaction
parameters are determined via calculation of second variation of total
energy for small deviation of magnetic moments \cite{Jref1}. This method
allows one to calculate $J(\mathbf{q})$, the lattice Fourier transform of
the exchange interactions $J(\mathbf{R}_{l})$. Thus one can easily calculate
long-range exchange interactions accurately even in complicated systems like
Cu$_{3}$TeO$_{6}$ here owing a highly-interconnected three-dimensional spin
network. Recently this technique has been used successfully for evaluating
magnetic interactions including DMIs in a series of materials \cite%
{ourJ1,Jref1,Jref3,Jref4,Jref5,Jref6,ourJ2,Jref7,Jref8,EuO}, and is employed
in this work to estimate the spin exchange parameters $J^{\prime }s$\ and DMIs \cite%
{ourJ1}.

\section{Results}

\begin{figure}[tbp]
\centering\includegraphics[width=0.45\textwidth]{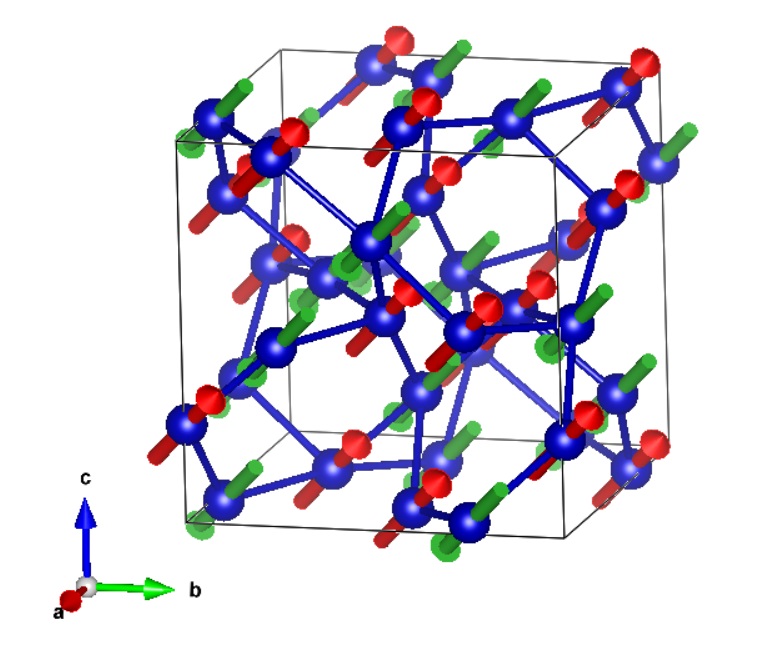}
\caption{Crystal structure of Cu$_{3}$TeO$_{6}$ \protect\cite{ctostr,ctostr2}%
. Only Cu ions are shown for simplicity. We refer to the spins as 
red and green arrows along [111] direction.}
\label{cry}
\end{figure}

\begin{figure}[tbp]
\centering\includegraphics[width=0.45\textwidth]{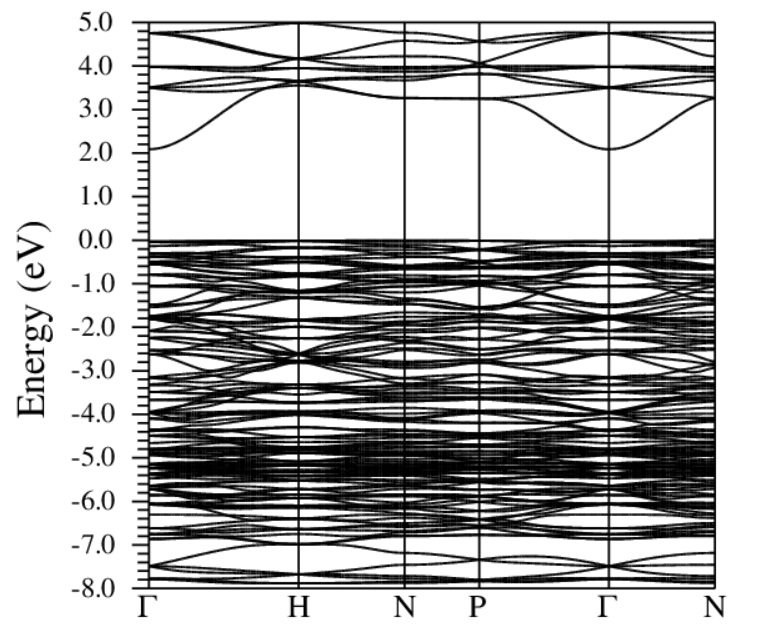}
\caption{Band structures of Cu$_{3}$TeO$_{6}$ from LSDA + $U$ (= 10 eV)
calculation with antiferromagnetic configuration. The Fermi energy is set to
zero.}
\label{bandstr}
\end{figure}

\begin{figure}[tbp]
\centering\includegraphics[width=0.45\textwidth]{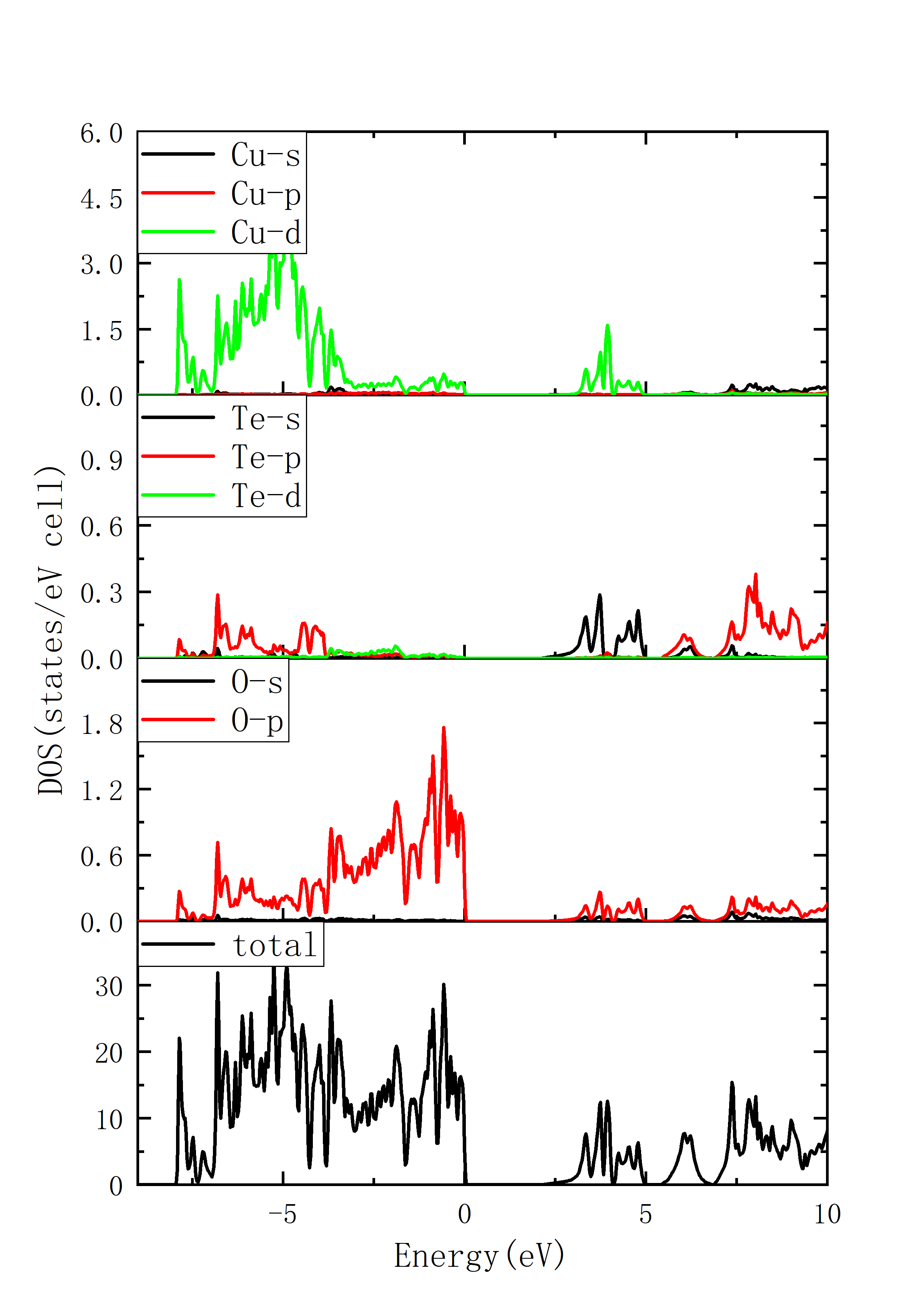}
\caption{Partial density of states (PDOS) of Cu$_{3}$TeO$_{6}$ from LSDA + $%
U $ (= 10 eV) calculation. The Fermi energy is set to zero. }
\label{dos}
\end{figure}

As shown in Fig. \ref{cry}, Cu$_{3}$TeO$_{6}$ crystallizes in the
centrosymmetric spin-web compound. The highly-interconnected
three-dimensional spin network consists of 12 Cu ions per primitive cell,
where six Cu ions form an almost coplanar hexagon and each Cu ion is shared
by two hexagons. Based on the collinear antiferromagnetic configuration
suggested by neutron powder diffraction experiment \cite{ctoexp-2} as shown
in Fig. \ref{cry}, we perform the first-principles calculations. Here we
adopt the LSDA + $U$ (= 10 eV) scheme, which is adequate for the
magnetically ordered insulating states \cite{RMP}. The band structures and
the density of states are shown in Fig. \ref{bandstr} and Fig. \ref{dos},
respectively. Our calculations indicate that Cu$_{3}$TeO$_{6}$ is an
insulator with a\ band gap about 2.07 eV from LSDA + $U$ (= 10 eV){\
calculations}. The O-$2p$ states are mainly located between $-$8.0 and 0.0
eV, while Te $5s$ and $5p$ bands appear mainly above 3.0 eV. Hence the
nominal valence for Te is $+6$\ while that for O is $-2$. As a result, Cu
ions have the nominal valence of +2, indicating the $3d^{9}$ electronic
configuration. The nine $3d$ occupied states of Cu ions are mainly located
from $-$8.0 to $-$3.0 eV, implying strong hybridization between Cu and O
states. Meanwhile the only one unoccupied state of Cu$^{2+}$ ions appears
mainly between 3.0 to 5.0 eV. Despite of strong hybridization between Cu and
O states, the calculated magnetic moment at the O site is negligible ($\sim $
0.01 $%
\mu
_{B}$), and the major magnetic moment is located at the Cu site. The
calculated magnetic moment of the Cu ions is 0.81 $%
\mu
_{B}$, which is smaller than the ideal $3d^{9}$ ($S$= $1/2$) configuration
and larger than the experimental value 0.64 $%
\mu
_{B}$ \cite{ctoexp-2}.

\begin{table}[tbp]
\caption{Calculated spin exchange parameters (in meV) evaluated from LSDA + $%
U$ (= 10 eV) scheme. The Cu-Cu distances and the corresponding number of
neighbours are presented in the 2nd the 3rd columns. The fitting spin
exchange parameters in the experimental work are also presented for
comparison.}
\label{ourJ}%
\begin{tabular}{llllll}
\hline\hline
& Distance(\AA ) & NN & Ref. \cite{ref3} & Ref. \cite{ref2} & Our results \\
\hline
$J_{1}$ & 3.18 & 4 & 9.07 & 4.49 & 7.05 \\
$J_{2}$ & 3.60 & 4 & 0.89 & -0.22 & 0.51 \\
$J_{3}$ & 4.77 & 2 & -1.81 & -1.49 & 0.04 \\
$J_{4}$ & 4.81 & 2 & 1.91 & 1.33 & 2.18 \\
$J_{5}$ & 4.81 & 2 & 1.91 & 1.79 & 0.09 \\
$J_{6}$ & 5.48 & 4 & 0.09 & -0.21 & 0.01 \\
$J_{7}$ & 5.73 & 4 & 1.83 & -0.14 & -0.01 \\
$J_{8}$ & 5.97 & 4 & -- & 0.11 & 0.04 \\
$J_{9}$ & 6.21 & 4 & -- & 4.51 & 3.77 \\
$J_{10}$ & 6.34 & 2 & -- & -- & 0.56 \\
$J_{11}$ & 6.34 & 2 & -- & -- & -0.01 \\
$J_{12}$ & 6.74 & 4 & -- & -- & 0.02 \\
$J_{13}$ & 7.17 & 2 & -- & -- & -0.06 \\
$J_{14}$ & 7.17 & 2 & -- & -- & -0.04 \\
$J_{15}$ & 7.27 & 4 & -- & -- & 0.00 \\
$J_{16}$ & 7.46 & 4 & -- & -- & 0.02 \\
$J_{17}$ & 7.64 & 4 & -- & -- & 0.10 \\
$J_{18}$ & 7.83 & 4 & -- & -- & 0.00 \\
$J_{19}$ & 8.26 & 4 & -- & -- & -0.02 \\
$J_{20}$ & 8.26 & 4 & -- & -- & 0.00 \\ \hline\hline
\end{tabular}%
\end{table}

\begin{figure}[tbp]
\centering\includegraphics[width=0.5\textwidth]{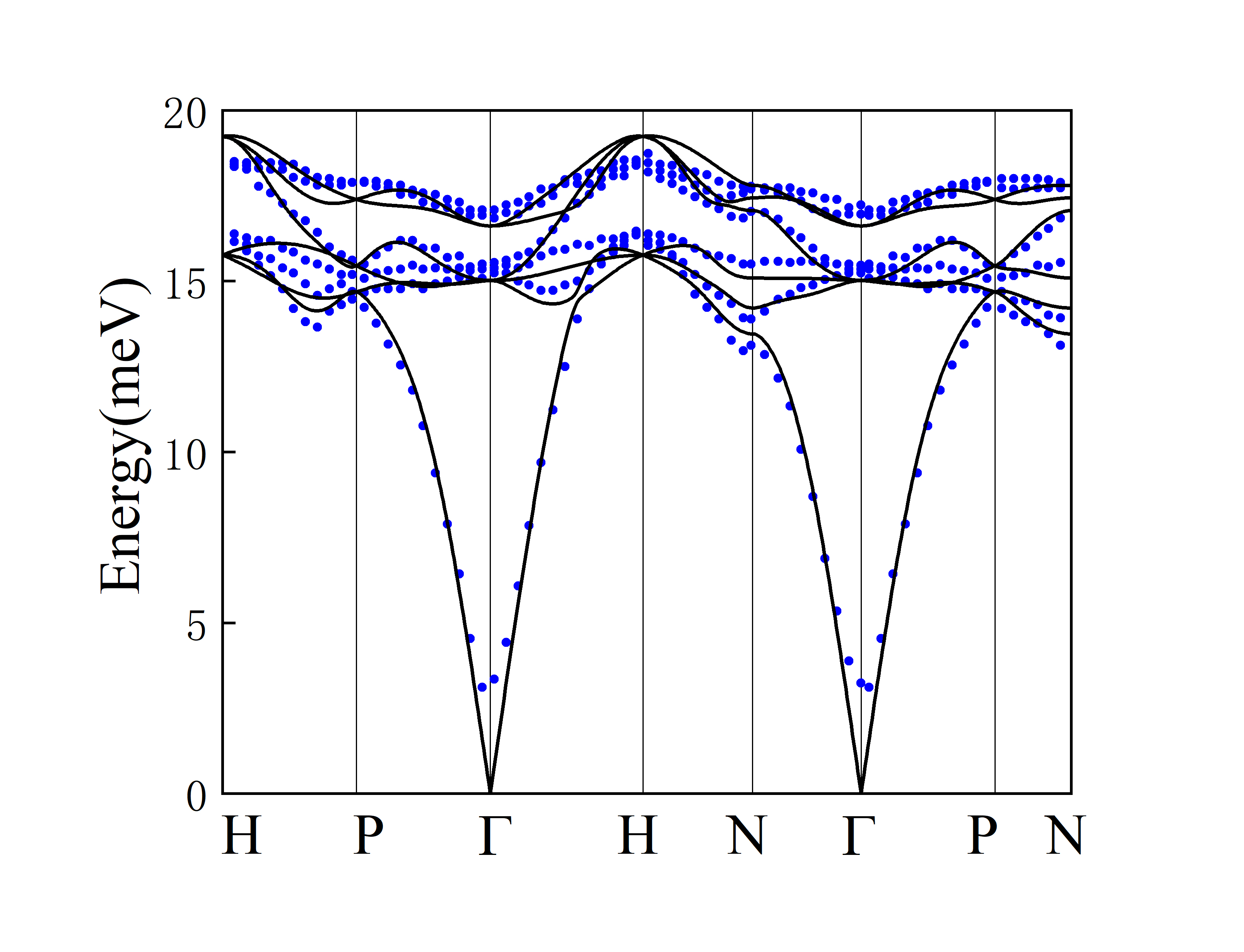}
\caption{Calculated spin-wave dispersion curves along high-symmetry axis for
Cu$_{3}$TeO$_{6}$. The INS spectra in Ref. \protect\cite{ref2} are also
shown as discrete points for comparison.}
\label{spinwave1}
\end{figure}

\begin{figure*}[tbp]
\centering\includegraphics[width=0.95\textwidth]{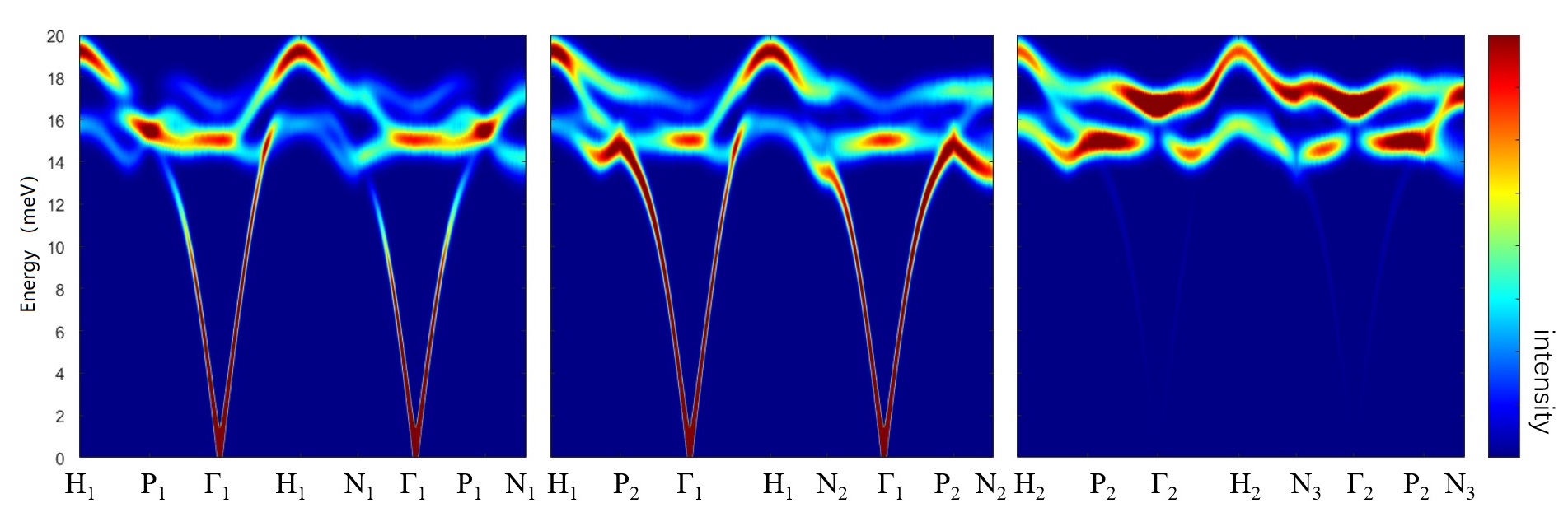}
\caption{Calculated magnon spectra for Cu$_{3}$TeO$_{6}$ \ along three
different momentum trajectories. The positions of high-symmetry points in
reciprocal lattice units are $H_{1}$(2,1,2), $H_{2}$(1,0,2), $P_{1}$($\frac{3%
}{2}$,$\frac{3}{2}$,$\frac{3}{2}$), $P_{2}$($\frac{3}{2}$,$\frac{1}{2}$,$%
\frac{3}{2}$), $\Gamma _{1}$(1,1,2), $\Gamma _{2}$(2,0,2), $N_{1}$($\frac{3}{%
2}$,$\frac{3}{2}$,2), $N_{2}$($\frac{3}{2}$,1,$\frac{3}{2}$), and $N_{3}$($%
\frac{3}{2}$,0,$\frac{3}{2}$), respectively.}
\label{spinwave_new2}
\end{figure*}

Based on the calculated electronic structures, we estimate the spin exchange
parameters $J^{\prime }s$\ (we refer to the exchange interaction of the $i$%
th NN as $J_{i}$) \cite{ourJ1}. The FPLR approach allows us to
calculate long-range exchange interactions accurately, and the results show
that these exchange parameters decrease rapidly with increasing distance. We
summarize the results up to 20th NN interaction $J_{20}$ in Table \ref{ourJ}%
. The fitted spin exchange parameters in the previous experimental work are
also shown for comparison \cite{ref2,ref3}. As the FPLR approach
automatically incorporates all the symmetry restrictions on exchange
interactions, we can distinguish the inequivalent $J^{\prime }$s even though
their exchange pathes own the same distance, such as $J_{4}$ and $J_{5}$
shown in the Table \ref{ourJ}. The results show that for the Cu-Cu bond with
the distance more than 7 \r{A}, the exchange interactions can be neglected
and there are only several sizable terms, including $J_{1}$, $J_{2}$, $J_{4}$%
, $J_{9}$, and $J_{10}$ . The strongest terms $J_{1}$ and $J_{9}$ both favor
antiferromagnetic ordering, which is compatible with the magnetic ground
state, thus there is no frustration between them. On the contrary the rest
three sizable terms $J_{2}$, $J_{4}$, and $J_{10}$ are not compatible with
the magnetic ground states. Note that they are much smaller than $J_{1}$ and
$J_{9}$ , which results in the modest frustration in Cu$_{3}$TeO$%
_{6}$ system, consistent with the experimental result \cite{ref3}. Based on
the obtained spin exchange parameters $J^{\prime }s$ as shown in the last
column in Table \ref{ourJ}, we calculate the Curie-Weiss temperature $\theta
_{CW}$ using the mean-field approximation theory \cite{theta} and the
calculated $\theta _{CW}$ is $-$147 K, comparable with several experimental
values as $-130$ K \cite{ctoexp-2}, $-165$ K \cite{ref2}\ and $-175$ K \cite%
{ref3}.

Using LSWT, we also calculate the magnetic excitation spectra and show the
spin-wave dispersion $\omega _{m}(\mathbf{q})$ ($m=1,2,\ldots ,12$) along
high-symmetry axis by solid lines in Fig. \ref{spinwave1}. For comparison
the INS spectra in the previous experimental work \cite{ref2} are also shown
as blue discrete points in Fig. \ref{spinwave1}. \footnote{%
The magnetic excitation spectra of the two experimental works \cite%
{ref2,ref3} are consistent with each other, but there is only one picture
containing discrete points which can be used for comparison in Ref. \cite%
{ref2}.} The calculated spin-wave dispersion is in well agreement with the
experimental measurements. As shown in Fig. \ref{cry}, there are 12 Cu ions
in each primitive cell, resulting in 12 bands in magnetic excitations. It
should be noted that, due to $\mathcal{P}\mathcal{T}$ symmetry, all the
magnon branches are doubly degenerate and one can only see six branches as
shown in Fig. \ref{spinwave1}. The acoustic branches extend up to about 15
meV, while the optical ones are mainly located between 15 and 20 meV. These
six doubly-degenerate branches form three Dirac points at $P$ at 14.7 meV,
15.4 meV and 17.4 meV, respectively, which is in good agreement with
experimental results of two points around 15 meV and one in 17.8 meV \cite%
{ref2}. Our numerical results show that there is a Dirac point at $\Gamma $
point of 16.6 meV while the experimental spin wave dispersions suggest the
point near 17 meV \cite{ref2}. In addition, we also reproduce a triply
degenerate node in 15.0 meV at $\Gamma $ point, as observed by Bao et al.
\cite{ref3} at the same energy of 15 meV. At $H$ point, we reproduce two
triply degenerate points are located at 15.8 meV and 19.2 meV, while the
experimental band crossings appear at about 16 meV and 18.5 meV \cite{ref3}.
Overall, these results are in good agreement with the theoretical and
experimental works \cite{ref1,ref2,ref3}.

Besides the spin wave dispersion, we also calculate the magnetic neutron
scattering intensity $I(\mathbf{Q,\omega })$ as a function of momentum $%
\mathbf{Q}=\mathbf{q}+\mathbf{G}$ ($\mathbf{G}$ is a reciprocal lattice
vector and $\mathbf{q}$ is in the 1st BZ) and energy $\omega $ by using
spin-spin correlation function, as shown in Fig. \ref{spinwave_new2}(a)$-$%
(c). Note that, the magnetic neutron scattering intensities may be
significantly different for different $\mathbf{G}$ at any giving $%
\mathbf{q}$. For comparison with the experiment \cite{ref2}, we display $I(%
\mathbf{Q,\omega })$ along three momentum trajectories shown in Figs. \ref%
{spinwave_new2}.\ The results capture most of the features in the previous
experiment \cite{ref2,ref3}. For example, the\ INS\ intensities in Figs. \ref%
{spinwave_new2}(a) and (b) are distributed in both acoustic and optical
branches, while the INS intensity in Fig. \ref{spinwave_new2}(c) is mainly
located at the optical branches between 15 and 20 meV. At $\Gamma _{1}$%
(1,1,2) point, the INS intensity is mainly located at acoustic branch and
the triply degenerate point, while the intensity at $\Gamma _{2}$(2,0,2)
point appears\ mainly at the Dirac point of 16.6 meV. Both at $H_{1}$(2,1,2)
and $H_{2}$(1,0,2) point, the intensity is mainly located at the branch of
the highest energy. As $\mathbf{Q}$ moves from $P_{2}$($\frac{3}{2}$,$\frac{1%
}{2}$,$\frac{3}{2}$) to $H_{1}$(2,1,2), the main intensity is located at the
lowest energy branch, as well as in the path $P_{2}(\frac{3}{2},\frac{1}{2},%
\frac{3}{2})-N_{2}$($\frac{3}{2}$,1,$\frac{3}{2}$). These results are
consistent with the experimental works \cite{ref2,ref3}.

In Ref. \cite{ref2}, through checking the magnon eigenvalue $\omega _{m}(%
\mathbf{q})$ at four high symmetry points, $\Gamma =(0,0,0),P=(\frac{1}{2},%
\frac{1}{2},\frac{1}{2}),H=(0,1,0),N=(\frac{1}{2},\frac{1}{2},0)$, Yao et al
obtained an interesting relation expressed by the following
eigenvalue-version of compatibility relation:
\begin{equation}
\sum_{m}(\omega _{\Gamma ,m}^{2}+4\omega _{P,m}^{2}+\omega
_{H,m}^{2})=\sum_{m}6\omega _{N,m}^{2},  \label{sumrule}
\end{equation}%
which they called \textquotedblleft sum rule\textquotedblright\ and proved
that it holds at least up to the 9th NN \cite{ref2}. Such kind of
sum rule is surprised to us for conventional compatibility relation is
usually only about the symmetry representations. Hence in the following we
analytically investigate whether there exists a general sum rule on
earth. We adopt the Heisenberg magnetic model as written by $\hat{H}=\frac{1%
}{2}\sum_{l,n,l^{\prime },n^{\prime }}J(\mathbf{R}_{l}+\boldsymbol{\tau }%
_{n},\mathbf{R}_{l}^{\prime }+\boldsymbol{\tau _{n^{\prime }}})\mathbf{S}%
_{ln}\cdot \mathbf{S}_{l^{\prime }n^{\prime }}$ where $l,l^{\prime }$ label
the unit cell and $n,n^{\prime }$ label the Cu ions: $n,n^{\prime
}=1,2,\ldots ,6$ represent Cu with up spin while $n,n^{\prime }=7,8,\ldots
,12$ represent Cu with down spin. The positions for these Cu ions in the
primitive unit cell are key for the following analysis, we thus
shown them in Tables \ref{taus1} of the Appendix. Based on the
antiferromagnetic ground state and using the LSWT, the magnon spectra are
obtained by diagonalizing the following matrix and then extracting the
non-negative eigenvalues for genuine magnon excitations \cite{spinwave} :
\begin{equation}
\mathcal{H}_{SW}(\mathbf{q})=\left[
\begin{array}{cc}
\mathcal{J}(\mathbf{q}) & \mathcal{J}^{\prime }(\mathbf{q}) \\
-\mathcal{J}^{\prime \dag }(\mathbf{q}) & -\mathcal{J}(-\mathbf{q})^{\top }%
\end{array}%
\right] ,  \label{Hsw}
\end{equation}%
where $\mathcal{J}(\mathbf{q})$ and $\mathcal{J}^{\prime }(\mathbf{q})$
(both are 12$\times $12 matrices) are expressed by:
\begin{widetext}
\begin{eqnarray}
 & \mathcal{J}(\mathbf{q})_{n,n'}=\zeta(n,n')\sum_{l}J(\boldsymbol{\tau}_n,\boldsymbol{\tau}_{n'}+\mathbf{R}_l)e^{i\mathbf{q}\cdot\mathbf{R}_l}+\delta_{n,n'}[\sum_{l,n''}J(\boldsymbol{\tau}_n,\boldsymbol{\tau}_{n''}+\mathbf{R}_l)
  -\sum_{l,n'''}J(\boldsymbol{\tau}_n,\boldsymbol{\tau}_{n'''}+\mathbf{R}_l)],\label{JqJq1}\\
 & \mathcal{J}'(\mathbf{q})_{n,n'}=\zeta'(n,n')\sum_{l}J(\boldsymbol{\tau}_n,\boldsymbol{\tau}_{n'}+\mathbf{R}_l)e^{i\mathbf{q}\cdot\mathbf{R}_l},\label{JqJq2}
\end{eqnarray}
where $\delta_{n,n'}$ is the Kronecker delta function, $n''$ runs through the Cu ions with spins parallel to that of the $n$th Cu while  $n'''$ runs through ions with spins antiparallel to that of the $n$th Cu.  $\zeta(n,n')$ ($\zeta'(n,n')$) is equal to 1 when the spins for the $n$th and $n'$th Cu's are parallel (antiparallel) otherwise equal to zero.
\end{widetext}N{ote that} $\sum_{m}\omega _{m}(\mathbf{q})^{2}=tr(\mathcal{H}%
_{SW}^{2})$, thus Eq. (\ref{sumrule}) can be written in the following form:
\begin{equation}
tr(\mathcal{H}_{SW}(\Gamma )^{2}+\mathcal{H}_{SW}(H)^{2}+4\mathcal{H}%
_{SW}(P)^{2}-6\mathcal{H}_{SW}(N)^{2})=0.  \label{sum}
\end{equation}

Firstly we consider the 1st NN. For Cu ion labelled by $n=1$, there are four
1st NNs, as shown in the first 4 rows of the Table \ref{1NN} of the Appendix. 
Cu ions in this compound occupy the 24$d$ Wyckoff
positions, and there are in total 24 Cu-Cu 1st NN bonds as also listed in
Table \ref{1NN} of the Appendix. We use $(n,n^{\prime },R_{l})$\
to denote the bond formed by the Cu ions labeled by $\tau _{n}$
and $\tau _{n^{\prime }}+R_{l}$. With these data, we can obtain all
the matrix elements of $H_{SW}$ for any given $\mathbf{q}$. For
each pair $(n,n^{\prime })$ of Cu ions, there is at most one
nearest-neighbor exchange path connecting them as shown in Table \ref{1NN}%
. For the mentioned four high symmetry points, the nondiagonal
matrix elements of $\mathcal{H}_{SW}$ are found to be one of the following
values $\pm J_{1}$, or $\pm iJ_{1}$, or 0. While the diagonal matrix
elements are equal to a constant for any $\mathbf{q}$. Therefore we
prove that $tr(\mathcal{H}_{SW}(\Gamma )^{2})=tr(\mathcal{H}%
_{SW}(H)^{2})=tr(\mathcal{H}_{P}(\Gamma )^{2})=tr(\mathcal{H}_{SW}(N)^{2})$
and Eq. (\ref{sum}) is satisfied for the 1st NN. Similarly, we can prove
that Eq. (\ref{sum}) holds from the $2$th NN to the $11$th NN. Further we
can prove that Eq. (\ref{sumrule}) holds with the exchanges up to 11th NNs%
. However for the 12th NN, as shown in Table \ref{12NN} of the
Appendix, for each pair $(n,n^{\prime })$, there may exist 4 exchange pathes
connecting them, so the corresponding matrix element of $\mathcal{H}%
_{SW}/J_{12}$ are the summation of four terms by Eqs. (\ref{JqJq1}) and (\ref%
{JqJq2}). This situation is different from that for the 1st NN, and one can
easily prove that Eq. (\ref{sum}) is no longer right for 12th NNs. Therefore
the \textquotedblleft sum rule\textquotedblright\ (i.e. Eq. (\ref{sumrule}%
))\ holds but only up to the 11th NNs.


It is worth mentioning that, though the Cu$_{3}$TeO$_{6}$ system has a
global inversion center, most of the Cu-Cu bonds don't own inversion
symmetry. Within the distance of 7 \r{A}, only the DMIs for 5th NN and 11th
NN are required to be vanishing because their bonds have inversion center.
Using the FPLR approach \cite{ourJ1}, we calculate the DMIs. Since the
strength of DMI is proportional to the corresponding $J$, we only calculate
the $D_{1}$ and $D_{9}$ (we refer to the DMI of the $i$th NN as $D_{i}$).
The $D_{1}$\ for the Cu-Cu bond between (0, 0.25, 0.969) and ($-$0.031, 0.5,
0.75) in the coordinate system is estimated to be (0.05, 0.25, 0.34) meV.
The direction of $D_{1}$ is nearly parallel to the normal direction of the
triangle formed by the three atoms in the Cu-O-Cu bond, which is consistent
with the physical expectation. While our calculation show that $%
D_{9}$ is very small ($\left\vert D_{9}\right\vert =0.06$ meV) and
have little effect on the magnetic configuration. Our numerical
ratio of $\left\vert D_{1}\right\vert /J_{1}$ is about 0.06, which is
smaller than the pure theoretical model estimation (0.2) \cite{ref1}. The
calculated DMIs result in a canting angle about 1.3$%
{{}^\circ}%
$, which is in agreement with the experimental value $\sim $6$%
{{}^\circ}%
$ \cite{ctoexp-2}. The size of the nodal line is proportional to square of
the ratio of DMI and $J$ \cite{ref1}, thus it is hard to identify
the nodal lines from the Dirac points for the current experiments.

\section{Conclusion}

In conclusion, using first-principles calculation, we presented a
comprehensive investigation of Cu$_{3}$TeO$_{6}$. The calculations show that
Cu$_{3}$TeO$_{6}$ is an insulator with a band gap about 2.07 eV and the
calculated magnetic moment of the Cu ions is 0.81 $%
\mu
_{B}$. Using magnetic force theorem and a first-principles linear-response
approach, we estimate the spin exchange parameters. The calculated exchange
parameters are short-range and can be neglected for the distance more than 7
\r{A}. The strongest terms $J_{1}$ and $J_{9}$ are compatible with the
magnetic ground state, while the terms $J_{2}$, $J_{4}$, and $J_{10}$ are
much smaller and not compatible with the magnetic ground states, which is
consistent with the modest frustration in this compound. We calculated the
magnon spectra using linear spin wave theory and the calculated spin wave is
in good agreement with the experiment. We also prove analytically that the
\textquotedblleft sum rule\textquotedblright\ proposed in Ref. \cite{ref2}
only holds up to the 11th nearest-neighbour interactions. The calculated
DMIs lead to a very small canting angle about 1.3$%
{{}^\circ}%
$ of non-collinear antiferromagnetic order. The weak DMIs are the possible
reason why the previous experimental work did not observe the nodal lines.

\section{Acknowledgement}

We wish to thank Prof. Yuan Li for discussion. The work was supported by
National Key R\&D Program of China (No. 2018YFA0305704 and 2017YFA0303203),
the NSFC (No. 11525417, 11834006, 51721001 and 11790311), the Fundamental
Research Funds for the Central Universities (No. 020414380085). DW was also
supported by the program B for Outstanding PhD candidate of Nanjing
University.

\clearpage
\setcounter{table}{0}

\section{Appendix}

In the Appendix, we list the coordinates of 12 Cu ions in Table \ref{taus1}.

\begin{table}[tbph]
\centering%
\begin{tabular}{|c|c|}
\hline
$n$ & $\boldsymbol{\tau }_{n}$ \\ \hline
1 & $(x,0,\frac{1}{4})$ \\ \hline
2 & $(\frac{1}{2}-x,0,\frac{3}{4})$ \\ \hline
3 & $(\frac{1}{4},x,0)$ \\ \hline
4 & $(\frac{3}{4},\frac{1}{2}-x,0)$ \\ \hline
5 & $(0,\frac{1}{4},x)$ \\ \hline
6 & $(0,\frac{3}{4},\frac{1}{2}-x)$ \\ \hline
7 & $(-x,0,-\frac{1}{4})$ \\ \hline
8 & $(\frac{1}{2}+x,0,\frac{1}{4})$ \\ \hline
9 & $(-\frac{1}{4},-x,0)$ \\ \hline
10 & $(\frac{1}{4},\frac{1}{2}+x,0)$ \\ \hline
11 & $(0,-\frac{1}{4},-x)$ \\ \hline
12 & $(0,\frac{1}{4},\frac{1}{2}+x)$ \\ \hline
\end{tabular}%
\caption{The coordinates of the 12 Cu ions in the conventional unit cell
basis vectors. $x=0.96907$.}
\label{taus1}
\end{table}

According to Eqs. (\ref{Hsw},\ref{JqJq1},\ref{JqJq2}) of the main
text, it is very easy to calculate the matrix $H_{SW}$  for any
wave vector $q$  when knowing the full information of all the
bonds.\ We thus give the detailed information of all the bonds connecting Cu
ions for the 1st NN and 12th NN in Tables \ref{1NN} and \ref{12NN},
respectively.
\begin{table}[tbph]
\centering%
\begin{tabular}{|c|c|c|}
\hline
$n$ & $n^{\prime }$ & $\mathbf{R}_{l}$ \\ \hline
1 & 9 & $(1,1,0)$ \\ \hline
1 & 10 & $(\frac{1}{2},-\frac{3}{2},\frac{1}{2})$ \\ \hline
1 & 11 & $(1,0,1)$ \\ \hline
1 & 12 & $(1,0,-1)$ \\ \hline
2 & 9 & $(0,1,1)$ \\ \hline
2 & 10 & $(\frac{1}{2},-\frac{3}{2},\frac{1}{2})$ \\ \hline
2 & 11 & $(-\frac{1}{2},\frac{1}{2},\frac{3}{2})$ \\ \hline
2 & 12 & $(-\frac{1}{2},-\frac{1}{2},-\frac{1}{2})$ \\ \hline
3 & 7 & $(1,1,0)$ \\ \hline
3 & 8 & $(-1,1,0)$ \\ \hline
3 & 11 & $(0,1,1)$ \\ \hline
3 & 12 & $(\frac{1}{2},\frac{1}{2},-\frac{3}{2})$ \\ \hline
4 & 7 & $(\frac{3}{2},-\frac{1}{2},\frac{1}{2})$ \\ \hline
4 & 8 & $(-\frac{1}{2},-\frac{1}{2},-\frac{1}{2})$ \\ \hline
4 & 11 & $(1,0,1)$ \\ \hline
4 & 12 & $(\frac{1}{2},-\frac{1}{2},-\frac{3}{2})$ \\ \hline
5 & 7 & $(1,0,1)$ \\ \hline
5 & 8 & $(-\frac{3}{2},-\frac{1}{2},\frac{1}{2})$ \\ \hline
5 & 9 & $(0,1,1)$ \\ \hline
5 & 10 & $(0,-1,1)$ \\ \hline
6 & 7 & $(1,1,0)$ \\ \hline
6 & 8 & $(-\frac{3}{2},\frac{1}{2},-\frac{1}{2})$ \\ \hline
6 & 9 & $(\frac{1}{2},\frac{3}{2},-\frac{1}{2})$ \\ \hline
6 & 10 & $(-\frac{1}{2},-\frac{1}{2},-\frac{1}{2})$ \\ \hline
\end{tabular}%
\caption{The 24 bonds for the first NN: each bond is characterized by the
positions of the two endings: $\boldsymbol{\protect\tau }_{n},\boldsymbol{%
\protect\tau }_{n^{\prime }}+\mathbf{R}_{l}$. }
\label{1NN}
\end{table}

\begin{table}[tbph]
\centering%
\begin{tabular}{|c|c|c|}
\hline
$n$ & $n^{\prime }$ & $\mathbf{R}_{l}$ \\ \hline
1 & 8 & $(-\frac{1}{2},-\frac{1}{2},-\frac{1}{2})$ \\ \hline
1 & 8 & $(-\frac{1}{2},\frac{1}{2},\frac{1}{2})$ \\ \hline
1 & 8 & $(-\frac{1}{2},-\frac{1}{2},\frac{1}{2})$ \\ \hline
1 & 8 & $(-\frac{1}{2},\frac{1}{2},-\frac{1}{2})$ \\ \hline
2 & 7 & $(\frac{1}{2},\frac{1}{2},\frac{1}{2})$ \\ \hline
2 & 7 & $(\frac{1}{2},-\frac{1}{2},\frac{3}{2})$ \\ \hline
2 & 7 & $(\frac{1}{2},-\frac{1}{2},\frac{1}{2})$ \\ \hline
2 & 7 & $(\frac{1}{2},\frac{1}{2},\frac{3}{2})$ \\ \hline
3 & 10 & $(-\frac{1}{2},-\frac{1}{2},-\frac{1}{2})$ \\ \hline
3 & 10 & $(\frac{1}{2},-\frac{1}{2},\frac{1}{2})$ \\ \hline
3 & 10 & $(-\frac{1}{2},-\frac{1}{2},\frac{1}{2})$ \\ \hline
3 & 10 & $(\frac{1}{2},-\frac{1}{2},-\frac{1}{2})$ \\ \hline
4 & 9 & $(\frac{1}{2},\frac{1}{2},\frac{1}{2})$ \\ \hline
4 & 9 & $(\frac{3}{2},\frac{1}{2},-\frac{1}{2})$ \\ \hline
4 & 9 & $(\frac{1}{2},\frac{1}{2},-\frac{1}{2})$ \\ \hline
4 & 9 & $(\frac{3}{2},\frac{1}{2},\frac{1}{2})$ \\ \hline
5 & 12 & $(-\frac{1}{2},-\frac{1}{2},-\frac{1}{2})$ \\ \hline
5 & 12 & $(\frac{1}{2},\frac{1}{2},-\frac{1}{2})$ \\ \hline
5 & 12 & $(\frac{1}{2},-\frac{1}{2},-\frac{1}{2})$ \\ \hline
5 & 12 & $(-\frac{1}{2},\frac{1}{2},-\frac{1}{2})$ \\ \hline
6 & 11 & $(\frac{1}{2},\frac{1}{2},\frac{1}{2})$ \\ \hline
6 & 11 & $(-\frac{1}{2},\frac{3}{2},\frac{1}{2})$ \\ \hline
6 & 11 & $(-\frac{1}{2},\frac{1}{2},\frac{1}{2})$ \\ \hline
6 & 11 & $(\frac{1}{2},\frac{3}{2},\frac{1}{2})$ \\ \hline
\end{tabular}%
\caption{The 24 bonds for the 12th NN: each bond is characterized by the
positions of the two endings: $\boldsymbol{\protect\tau }_{n},\boldsymbol{%
\protect\tau }_{n^{\prime }}+\mathbf{R}_{l}$ }
\label{12NN}
\end{table}

\clearpage
\bibliography{CTO_v7}

\end{document}